\documentclass{iaus}
\usepackage{graphicx}

\title[Star Forming Galaxies at $z >$ 5] 
{Star Forming Galaxies at $z >$ 5}

\author[Yoshi Taniguchi]   
{Yoshi Taniguchi}

\affiliation{Research Center for Space and Cosmic Evolution,
Ehime University, 2-5 Bunkyo-cho, Matsuyama 790-8577, Japan
\\email: {\tt tani@cosmos.ehime-u.ac.jp}}

\pubyear{2008}
\volume{250}  
\pagerange{000--000}
\jname{Massive Stars as Cosmic Engines}
\editors{A.C. Editor, B.D. Editor \& C.E. Editor, eds.}
\begin{document}

\maketitle

\begin{abstract}
We present recent progress in searching for galaxies at redshift from 
$z \simeq 5$ to $z \simeq 10$. Wide-field and senstive surveys with
8m class telescopes have been providing more than
several hundreds of star forming galaxies at $z \simeq$ 5 -- 7
that are probed in the optical window.
These galaxies are used to study the early cosmic star formation 
activity as well as the early structure formation in the universe.
Moreover, near infrared deep imaging and spectropscopic surveys 
have found probable candidates of galaxies from $z \simeq$ 7 to
$z \simeq 10$. Although these candidates are too faint to be identified 
unambiguously, we human being
are now going to the universe beyond 13 billion light years, close 
to the epoch of first-generations stars; i.e., Population III stars.
We also mention about challanges to find Population III-dominated 
galaxies in the early universe.  
\end{abstract}

\firstsection 
\section{Introduction}

Massive stars are key ingredients in the early phase of galaxy formation.
First stras (i.e., Popoulation III stars) are considered to be very massive from
theoretical aspects although the formation of low mass stras may not be ruled out.
Once first massive stars were formed in subgalactic gas clumps within dark matter halos,
they could begin to work as cosmic engines because of thier huge number of ionizing photons.
Then, since they could die after a few million years after their birth,
they also supply both kinetic energy and heavy elements even into the
intergalactic space. Therefore, massive stars also work as mechanical and 
chemical engines in the universe. 
All these issues suggest that 
the formation and early evolution of galaxies could be controlled by
massive stars; how massive, how many massive stars (i.e., more generally, 
the initial mass function), and when did they begin to form ?
It is also worthwhile noting that massive stars in the phase of galaxy formation
are considered to be related to the cosmic re-ionization of the universe.

In order to investigate these problems, it is necessary to carry out
searches for galaxies at very high redshift. 
Such high-$z$ galaxies have been identified mainly by the following survey methods;
(1) optical narrow-band surveys for strong Ly$\alpha$ emitters (LAEs), and (2)
optical broad-band surveys for Lyman break galaxies (LBGs) (i.e., the dropout technique;
e.g., Steidel et al. 1999, 2003; Ouchi et al. 2004).
Other methods are summarized in Taniguchi et al. (2003).
Although both methods need follow-up optical spectroscopy to confirm that 
photometrically selected objects are real high-$z$ galaxies, 
more than several hundreds of 
galaxies beyond $z$ = 5 have been already found to date thanks to the great observational
capability of 8m class telescope facilities
(e.g., Dey et al. 1998; Rhodes et al. 2003; Hu et al. 2002, 2004; Ajiki et al. 2003; Kodaira et al 2003;
Taniguchi et al. 2005; Kashikawa et al. 2006; Ouchi et al. 2008).
To probe massive star formation or initial starbursts in early universe, 
hydrogen Ly$\alpha$ emission provides a powerful tool
(Partridge and Peebles 1967; Haiman 2002).
Therefore, in this review, we discuss the observational nature 
star-forming galaxies at high redshift found mainly by the optical narrow-band 
imaging surveys made so far (section 2).

In section 3, we give a brief summary of near infrared deep surveys for
forming galaxies beyond $z = 7$ and discuss thier implication.
Another intersting issue is to search for evidence for Population III stars
in early universe. Since theoretical consideration suggests that some galaxies
at $z \gtrsim 2$  may be dominated by Population III stars because of low
feedback from first-generation supernovae. Therefore, in section 4, we give a summary
of recent trials for searching for such Population III-dominated galaxies. 
Finally, in section 5, we discuss future prospects in this research field briefly. 

\section{Surveys in the Optical Window; Star-forming galaxies at $z$ = 5 -- 7}

When we use the optical window (e.g., 400nm -- 1000nm), we are able to 
search for LAEs at $z \sim$ 2.5 -- 7. If we are interested in LAEs beyond 
$z =5$, surveys should be made at wavelengths longer than 700 nm where
OH airglow emission lines are bright. Therefore, we have to use some
narrow windows where OH airglow is moderately weak; e.g., 815nm, 920nm,
and so on. These atomspheric windows allow us to search for LAEs at $z \simeq 5.7$ and
$z \simeq 6.6$, and so on. A summary of such LAE surveys for $z > 5$ is given in Table 1.
We also give a list of the top ten of most distant galaxies found in the optical
LAE surveys in Table 2; the most distant LAE known to date is IOK1 at $z = 6.96$
(Iye et al. 2006). Note that candidate galaxies around $z \sim 7$ have been found in
the GLARE (Gemini Lyman-Alpha at Reionization Era) survey (Stanway et al. 2007).

We summarize the observational properties of LAEs at $z$ = 5 -- 7 as follows. 
(1) Stellar masses; $\sim 10^{9 - 10} M_{\odot}$, 
(2) Sizes (Ly$\alpha$ emission)); a few kpc,
(3) Ly$\alpha$ luminosity; $L$(Ly$\alpha$) $\sim 10^{42 - 43}$ erg s$^{-1}$;
(4) Stellar ages; several million to several hundreds million years;
(5) Star formation rates; a few to several tens $M_{\odot}$ y$^{-1}$; 
(6) Star formation rate densities; $\sim 10^{-3} ~  M_{\odot}$ y$^{-1}$ Mpc$^{-3}$;
(7) Morphology (Ly$\alpha$ emission); spatially extended, but only few information;
and
(8) Morphology (UV continuum); spatially extended (smaller than Ly$\alpha$ emission), 
but only few information (e.g., Rhoads et al. 2005; Taniguchi et al. 2008).

Note, however, that more massive galaxies tend to be found in high-$z$ LBG samples
with stellar masses of $\sim 10^{10 - 11} M_{\odot}$ (e.g., Egami et al. 2005;
Mobasher et al. 2005; Yan et al. 2006; Eyles et al. 2007 and references therein). 
Moreover, the star formation rate densities for high-$z$ LBGs is $\sim 10^{-2} ~  
M_{\odot}$ y$^{-1}$ Mpc$^{-3}$, being higher by one order of magnitude than those of
LAEs at similar redshifts (e.g., Taniguchi et al. 2005 and referernces therein).

The Ly$\alpha$ luminosity function for LAEs at $z > 5$ has been obtained for 
several samples of LAEs (e.g., Ajiki et al. 2004; Hu et al. 2004; 
Shimasaku et al. 2006; Kashikawa et al. 2006).
One interesting point is that bright LAEs appear to be rarer at $z \approx 6.5$
than  at $z \approx 5.7$ in the Subaru Deep Field (Kashikawa et al. 2006),
suggesting a possible evolutionary effect from $z \approx 6.5$ to $z \approx 5.7$.

As for large-scale structures in early universe, there are lines of evidence for
galaxy clustering below $z \approx 6$ [at $z \approx 5.7$ (Hu et al. 2004; Ouchi et al. 2005),
and at $z \approx 4.9$ (Shimasaku et al. 2003)] although there is no strong 
evidence for such clustering at $z \approx 5.7$ (Ajiki et al. 2003; Murayama et al. 2007).
Beyond $z=6$, no such clustering feature has been found to date (Kashikawa et al. 2006). 
These observations suggest that the large-scale structure probed by LAEs
could grow up from $z \sim 6$.

\begin{table}
  \begin{center}
  \caption{Top ten of high-redshift galaxies found in the optical surveys}
  \label{tab2}
 {\scriptsize
  \begin{tabular}{|c|c|c|c|c|c|c|}\hline 
Window & Field & {\it z} & {\it V} & $N$(photo) & $N$(sp) & Ref. \\  
 (nm) &  &  & (Mpc$^3$)  &  &  &   \\  \hline
973 & SDF & 6.96 & $3.2 \times 10^5$  & 2 & 1 & a \\ \hline
921 & SDF & 6.5 -- 6.6 & $3.2 \times 10^5$  & 58 & 34 & b, c, d \\ 
912 & 6 fields & 6.56 & --  & 1 & 1 & e \\ \hline
816 & SDSS1044 & 5.7 & $2.0 \times 10^5$  & 20 & 2 & f \\ 
815, 823 & LALA & 5.7 -- 5.8 & $\sim 2 \times 10^5$  & 18 & 2 & g \\
816 & SSA22 & 5.7 & $1.9 \times 10^5$  & 26 & 19 & h \\ 
816 & SDF & 5.7 & $1.9 \times 10^5$  & 89 & 37 & i \\ 
816 & GOODS-N & 5.7 & $4.0 \times 10^4$  & 10 & -- & j \\ 
816 & GOODS-S & 5.7 & $4.0 \times 10^4$  & 4 & -- & j \\ 
816 & COSMOS & 5.7 & $1.8 \times 10^6$  & 119 & -- & k \\ 
816 & SXDS & 5.7 & $9.2 \times 10^5$  & 401 & 17 & l \\ \hline
  \end{tabular}
  }
 \end{center}
\vspace{1mm}
 \scriptsize{
 {\it Field:}  SDF = Subaru Deep Field (Kashikawa et al. 2004), SDSS1044 = SDSSp J104433.04-012522.2
 (Fan et al. 2000), LALA = Large Area Lyman Alpha survey, SSA22 = Samll Selected Area at RA = 22h, 
 GOODS = Great Observatory Origins Deep Survey - North and South (Giavalisco et al. 2004),
 COSMOS = Cosmic Evolution Survey (Scoville et al. 2007), and SXDS = Subaru XMM-Newton
 Deep Survey (Sekiguchi et al. 2004) \\
 {\it Survey Volume ($V$):} A flat universe with $\Omega_{\rm m} = 0.3$,  $\Omega_\Lambda = 0.7$, and
  $H_0 = 70$ km s$^{-1}$ Mpc$^{-1}$ is adopted. \\ 
 {$N$(photo):} The number of LAEs identified by photometric conditions. \\
 {$N$(sp):} The number of LAEs confirmed by follow-up spectroscopy. Note that
  follow-up spectroscopy is going on in some cases. \\
 {\it References:}  a = Iye et al. 2006, Ota et al. 2008; b = Kodaira et al. 2003; c = Taniguchi et al. 2005;
  d = Kashikawa et al. 2006; e = Hu et al. 2002; f = Ajiki et al. 2003 (see also Ajiki
  et al. 2004); g = Rhodes \&
  Malhotra 2001, 2003; h = Hu et al. 2004; i =  Shimasaku et al. 2005; j =Ajiki et al. 2006;
  k = Murayama et al. 2007; l = Ouchi et al. 2008}
\end{table}

\begin{table}
  \begin{center}
  \caption{Top ten of high-redshift galaxies found in the optical surveys}
  \label{tab2}
 {\scriptsize
  \begin{tabular}{|c|c|c|c|}\hline 
{\bf No.} & {\bf Name} & {\bf Redshift} & {\bf Ref.} \\  \hline
1 & IOK1      & 6.96 & a  \\ \hline
2 & SDF132522.3+273520 & 6.597 & b  \\ \hline
3 & SDF132520.4+273459 & 6.596 & c  \\ \hline
4 & SDF132357.1+272446 & 6.589 & c  \\ \hline
5 & SDF132432.5+271647 & 6.580 & b  \\ \hline
6 & SDF132528.8+273043 & 6.578 & b  \\ \hline
6 & SDF132418.3+271455 & 6.578 & b, d  \\ \hline
8 & HCM-6A    & 6.56  & e  \\ \hline
9 & SDF132432.9+273124 & 6.557 & c  \\ \hline
10 & SDF132408.3+271543 & 6.554 & b  \\ \hline
  \end{tabular}
  }
 \end{center}
\vspace{1mm}
 \scriptsize{
 {\it References:}  a = Iye et al. 2006, Ota et al. 2008; 
  b = Taniguchi et al. 2005; c = Kashikawa et al. 2006;
  d = Kodaira et al. 2003; e = Hu et al. 2002}
\end{table}

\section{Surveys in the Near-Infrared Window; Star-forming galaxies at $z >$ 7}

The success of optical searches for high-$z$ galaxies up to $z \approx 7$ urged us
to carry out near-infrared (NIR) surveys for very high-$z$ galaxies beyond $z =7$.
Such galaxies, if any, could be much fainter than galaxies with $z < 7$ because they 
could be in still in a mass assmbly phase in their evolution. Moreover, the attenuation 
of H {\sc i} atoms should be much severe for detecting Ly$\alpha$ emission 
although cosmological H {\sc ii} regions around LAEs could help for such detection
(e.g., Haiman 2002). In this section, we give a summary of recent NIR surveys for 
forming galaxies.

\vspace{5mm}

[1] {\it Narrow-band Imaging Surveys}: The narrow-band imaging technique provides us 
a larage number of LAEs at $2 < z < 7$. Therefore, this technique should be applied 
in the NIR window.  

The first NIR trial was reported by Willis \& Courbin (2005). Their project name is
^^ ^^ ZEN (z equal nine)". They used a narroband filter centered at 1.187 $\mu$m
in $J$ band with VLT/ISAAC (FOV = 2.5 arcmin $\times$ 2.5 arcmin).  
They found no LAE candidate with $L$(Ly$\alpha$) $> 10^{43} ~ h^{-2}$ erg s$^{-1}$
in the survey volume of 340 $h^{-3}$ Mpc$^{-3}$ where $h = H_0 / 100$. 
They also made a similar deep NIR narroband imaging survey for LAEs in three
lensing cluster areas (Abell 114, 1689, and 1835). Again, they found no LAE candidate with 
$L$(Ly$\alpha$) $> 10^{43} ~ h^{-2}$ erg s$^{-1}$ 
in the survey volume of 1270 $h^{-3}$ Mpc$^{-3}$.

Using the same instrument and the NIR narrowband filter used in the ZEN survey,
Cuby et al. (2007) made a survey for LAEs at $z \approx 9$ in seven fields,
covering 31 arcmin$^2$ in total. Although their survey area is wider than that of ZEN,
they found no LAE candidate. 

\vspace{2mm}

[2] {\it Broad-band Imaging Surveys}: The broad-band imaging technique is also
powerful to detect high-$z$ galaxies. This is the same technique as the Lyman
break method in the optical. The deepest NIR imaging survey data are provided 
by the Hubble Ultra Deep Field (Bouwens et al. 2005; see also Bouwens \& 
Illingworth 2006). The survey depth is down to 
28.6 in $J_{110}$(AB) and 28.5 in $H_{160}$(AB) with a 0.6 arcsec aperture.
They found three probable galaxies at $z \sim 10$ with $H_{160} \simeq
28$, corresoponding to $\approx 0.3 L^*_{z=3}$. 
Unfortunately, all these sources are too faint to be observed in spectroscopy.
The presence of these objects gives a star formation rate density of 
$\sim 10^{-3}$ $M_{\odot}$ y$^{-1}$ Mpc$^{-3}$, being smaller by
one order of magnitude than that at $z$ = 5 -- 6.

When we use ground-based telescopes, the survey depth is around 25 AB magnitude
in $J$ and $H$ at best because of strong airglow emission. Therefore, 
NIR broadband imaging surveys for a blank field from the ground may not work well.
Or, we need help of gravitational lensing. Richard et al. (2008) made deep NIR
imaging survey for two lensing cluster regions (Abell 1835 and AC 114)
by using VLT/ISAAC. Their survey depth is 25.5 -- 25.6in $J$(AB), 24.7 in $H$(AB),
and 24.3 -- 24.7 in $K_{\rm s}$(AB) for the two regions.
They found 8 and 5 probabale candidates at $7 < z < 10$ in Abell 1835 and AC 114,
respectively. Given the magnification factor from 1.5 to 10, their star formation
rate ranges from a few to 20 $M_{\odot}$ y$^{-1}$. 

\vspace{2mm}

[3] {\it Blind Spectrocsopic Surveys along Caustic Lines}: This technique is
very unique but very powerful in finding intrinsically faint objects at
high redshift; see for optical trials, Ellis et al. (2001) and Santos 
et al. (2004).  Stark et al. (2007) made NIR spectrocopic surveys for 
gravitationally-lensed LAEs around nine interemedaite-$z$ clusters of 
galaxies by using Keck/NIRSPEC. Since their spectroscopic survey is designed to
sweep the caustic lines of the gravitational lensing, the amiplification factor
is as high as 10 -- 50 for objects around $z = 10$. They found 6 probable LAE
candidates at $z$ = 8.7 -- 10.2. Among them, two objects are the most likely 
candidates from their careful follow-up observations; Abell 68 c1 at $z = 9.32$
and Abell 2219 c1 at $z = 8.99$.

\vspace{5mm}

We have introduced five independent deep NIR surveys for high-$z$ galaxies
beyond $z =7$. Generally speaking, it is more difficult to find candidates
of forming galaxies in NIR than in optical.
One reason is that there is no very wide-field NIR imager on
8m class telescopes. MOIRCS on the Subaru Telescope has FOV = 4 arcmin $\times$ 7 arcmin
(http://www.subarutelescope.org/Observing/Instruments/MOIRCS/index.html)
and HAWK-I on VLT has FOV = 7 arcmin $\times$ 7 arcmin
(http://www.eso.org/instruments/hawki/). Although these NIR cameras
have widest FOV on such 8m class telescopes, their FOVs are still much smaller than 
those of optical cameras ($>$ 30 arcmin $\times$ 30 arcmin). Note that both the stronger 
airglow emission and the much severe Tolman dimming [i.e., the surface brightness
dimming as $(1+z)^{-4}$] also makes it difficult to detect LAEs at NIR.
Also, forming galaxies at $z > 7$ could be faint intrinsically.

\section{Surveys for Population III--dominated Galaxies}

One intersting issue is to address on first stras in our universe:
When were they made ? How massive were they ? How did they affect 
the nature of intergalactic medium, chemically and dynamically ?
Did only first stars contribute to the cosmic re-ionization ?
How were they related to the formation of supermassive black holes ? 
In order to give firm answers to these questions, it is necessary to
probe first stars or forming galaxies at very high redshift.

The major epoch of Pop III star formation may be at $z \sim$ 10 -- 30,
when dark matter halos with mass of $M_{\rm halo} \sim 10^6~ M_{\odot}$ could 
have grown up to form first stars
(see for a review, Loeb \& Barkana 2001).
Shortly after the formation of Pop III stars ($\sim 10^6$ years after),
first sypernova explosions could occur and then polute the gas inside and outside
of dark matter halos. However, if this feedback process is inefficient, 
Pop III stars could be born in galaxies even at $z \sim 2$ or so
(e.g., Scannapieco et al. 2003; Schneider et al. 2006; Jimenez \& Haiman 2006;
Tornatore et al. 2007; see also, for recent review, Norman 2008). 
If this is the case, it is possible to probe Pop III
dominated forming galaxies in the optical window. 

There are two important observational properties of such Pop III dominated 
galaxies. One is the large equivalent width (EW) of hydrogen Ly$\alpha$ emission.
The other property is moderately strong He {\sc ii} $\lambda$1640 emission.
These properties are attributed to the very high effective temperature 
($\sim 10^5$ K) of Pop III stars (e.g., Tumlinson \& Shull 2000; Tumlinson 
et al. 2001, 2003; Bromm et al. 2001; Oh et al. 2001; Schaerer 2002, 2003).
Interestingly, it has been reported that some high-$z$ LAEs have a large 
rest-frame equivalent width of Ly$\alpha$ emission with $EW_0 >$ a few 100 \AA ~
(e.g., Malhotra \& Rhoads 2002; Nagao et al. 2004, 2005a, 2007; Shimasaku et al. 
2006; Dijkstra \& Wyithe 2007). These observations may not be explained with
photoiniziation by usual massive stars. 

For one of such LAEs with a large EW Ly$\alpha$ emission, Nagao et al. (2005a)
made a ultra deep NIR spectroscopy to detect He {\sc ii} emission; 
SDF 132440.6+273607 with $EW_0$(Ly$\alpha$) = 130 \AA ~ and $L$(Ly$\alpha$) =
$1.8 \times 10^{43}$ erg s$^{-1}$ at $z = 6.33$. 
However, they found no He {\sc ii} feature, suggesting $L$(HeII) $< 1.4 \times 10^{42}$
erg s$^{-1}$. Stacking analyses of spectra of high-$z$ galaxies also found no evidence
for Pop III-driven He {\sc ii} emission; (1) Shapley et al. (2003) examined the stacked 
spectrum of $\simeq$ 1000 LBGs at $z \simeq$ 3 and obtained evidence for He {\sc ii}
emission. However, the authors suggested that this feaure may be attributed to 
Pop I hot stars such as WR stars (see also the comment given by Max Peetini
in the discussion; cf. Jimenez \& Haiman 2006). (2) Ouchi et al. (2008)
examined the stacked spectrum of $\simeq$ 50 LBGs at $z =3.7$ and found no He {\sc ii} 
emission. (3) See the comment given by Jon Eldridge in discussion.

Recently, Nagao et al. (2008) made a unique survey for Ly$\alpha$-He {\sc ii} emitters
by using combination of intermediate- and narrow-band filters in the optical window.
They used (i) IA598 and NB816 filters for LAEs at $z \approx 4.0$ and (ii) IA 679 and
NB921 for LAEs at $z \approx 4.6$, where the Subaru IA filter system consists of 20 filters
covering from 4000 \AA ~ to 9500 \AA ~ with a spectroscopi resolution of 
$R = \lambda/\Delta\lambda = 23$ (Taniguchi 2001; see also Ajiki et al. 2004; Yamada et al. 2005). 
Their survey field is SDF. Although they found 10 dual-line
emitters, they are not identified as Ly$\alpha$/He {\sc ii} emitters (i.e., low-$z$
dual-line emitters such as [O {\sc ii}]/[O {\sc iii}] ones.
Thier survey shows that there is no LAEs with Pop III star formation rate with
$> 2 M_{\odot}$ y$^{-1}$ and the Pop III star formation rate density is 
lower than $5 \times 10^{-6} M_{\odot}$ y$^{-1}$ Mpc$^{-3}$, being smaller by a few
orders of magnitudes than those of LAEs at similar redshifts.    

\section{Future Prospects}

Finally, we give comments on future prospects.
As shown in previous sections, optical deep surveys have been finding a large number of
forming galaxies at $z \sim$ 5 -- 7. However, NIR deep surveys have been facing to the 
technical limit of the exsiting large telescope facilities including the Hubble Space
Telescope. Therefore, we will have to wait for next-generation telescopes in order to 
improve our knowledge on massive stars at very high redshift (i.e., Pop III stars) and
stat formation hitory of in very young galaxies. 

We will have two types of new-generation telescopes in neaer future; one is
the James Weeb Space Telescope (JWST: Sonneborn 2008, see http://www.jwst.nasa.gov/)
and the other is extreme large telescopes (ELTs) on the ground 
such as the European-ELT (D'Odorico 2008; see http://www.eso.org/sci/facilities/eelt/) 
and the Thirty Meter Telescope (TMT: http://www.tmt.org/) led by Caltech and University of
California. Extremly high observational capabilities of these telescopes will open the door
to Pop III dominated universe. In particular, mid infrared observations with JWST
will be able to go to $z \sim 30$.

Yet, we will have to do our best to find much younger, Pop III-dominated  galaxies
at high redshift up to $z \sim 30$ before next-generation, space and ground-based
telescopes will come. 
For this effort, we have two nice helpers. One is the graviational lensing,
as demonstrated in section 3 (e.g., Stark et al. 2007; Richard et al. 2008).
The other helper is very bright gamma-ray bursts at high redshift. 
After the gamma ray bursts (GRBs), very bright optical flashes have been often observed.
The most distant GRB event known to date was found at $z = 6.33$ (Kawai et al. 2006).  
If such GRB events will occur at $z \sim 7$, we will be detect their rest-frame optical
flashes in the NIR windows (sse Fynbo 2008). 

As for the formation and evolution of galaxies, next-generation telescopes will allow us
to explore what happened in galaxy formation and in early evolution phase.
Then, we will be able to obtain a unified picture for various types of galaxies 
at high redshift, e.g., LAEs, LBGs, EROs (extremely red objects), DRGs (distant red 
galaxies), BzKs (galaxies selected from BzK photometry), SMGs (submillimeter galaxies),
and so on. Finally, we will learn how massive stars have been working as cosmic engines
for more than 10 billion years.

\vspace{5mm}

The author would like to thank all the members of both SOC and LOC of this wonderful
symposium at Kauai/Hawaii. He also thanks his colleagues, in particular, 
Nobunari Kashikawa and Tohru Nagao.

\begin{discussion}

\discuss{Eldridge}{We (me and Elizabeth Stanway) also have looked for He {\scshape ii} at high redshift.
We have a stacked spectrum of 50 galaxies with a mean redshift of 4.7 and 
we see no He {\scshape ii} to a quite low limit.}

\discuss{Taniguchi}{Thank you for your comment.}

\discuss{Pettini}{In our survey of UV-bright galaxies
at $z$ = 2 - 3, we have examples of galaxies with stellar
populations which seem very young, with ages of only
a few tens of million of years. These objects may be
the lower redshift analogues of your galaxies at
$z$ = 6 - 7 and, if they are, they are of course much
easier to study spectroscopically at the lower redshifts.
Somewhat surprising perhaps, even these very young
galaxies have already fairly high metallicities,
between 1/5 and 1/3 of solar. It seems that once
star-formation starts in these kinds of high redshift
galaxies, it proceeds very fast, so that the traces
of the first few generations of stars are quickly
swamped. Thus, you may find that it is also very
difficult to catch one of your z = 6 - 7 galaxies,
which are also undergoing vigorous star formation
(otherwise you would not see them), at such an early stage
that the signatures of Pop III stars can be discerned.}

\discuss{Taniguchi}{Thank you for your comment.}

\discuss{Taniguchi}{Now, one of the most important issues in this research field is 
to search for galaxies with Pop III stars at high redshift. Aslthough theoretical
considereations suggest that such galaxies could be found at $z > 2.5$.
However, any trials made so far failed to identify such galaxies.
I think that we need more systematic search for Pop III dominated galaxies 
at $z > 2.5$ by using a large sample of galaxies (i.e., LAEs) in near future. }

\end{discussion}


\begin{thebibliography}{}

\bibitem[]{}
Ajiki, M., et al. 2003,
\textit{AJ}, 126, 2091

\bibitem[]{}
Ajiki, M., et al. 2004,
\textit{PASJ}, 56, 597

\bibitem[]{}
Ajiki, M., et al. 2006,
\textit{ApJ}, 638, 596

\bibitem[]{}
Bouwens, R. J., Illingworth, G. D., Thompson, R. I., \& Franx, M. 2005
\textit{ApJ}, 624, L5

\bibitem[]{}
Bouwens, R. J., \& Illingworth, G. D. 2006
\textit{Nature}, 443, 189

\bibitem[]{}
Bromm, V., Kudritzki, R. P., \& Loeb, A. 2001,
\textit{ApJ}, 552, 464

\bibitem[]{}
Cuby, J. G., Hibon, P., Le Febre, O., Gilmozzi, R., Moorwood, A., \& 
ven der Werf, P. 2007,
\textit{AA}, 461, 911

\bibitem[]{}
Dey, A., Spinrad, H., Stern, D., Graham, J. R., \& Shaffee, F. H. 1998,
\textit{ApJ}, 498, L93

\bibitem[]{}
Dijkstra, M., \& Wyithe, J. S. B. 2006,
\textit{MNRAS}, 379, 1589

\bibitem[]{}
D'Odorico, D.  2008,
\textit{in this volume}

\bibitem[]{}
Egami, E., et al. 2005,
\textit{ApJ}, 618, L5


\bibitem[]{}
Ellis, R. S., Santos, M. R., Kneib, J. -P., \& Kuijken, K. 2001,
\textit{ApJ}, 560, L119

\bibitem[]{}
Eyles, L. P., Bunker, A. J., Ellis, R. S., Lacy, M., Stanway, E. R., 
Starck, D. P., \& Chiu, K. 2008,
\textit{MNRAS}, 374, 910


\bibitem[]{}
Fan, X., et al. 2000,
\textit{AJ}, 120, 1167

\bibitem[]{}
Fynbo, J. 2008,
\textit{in this volume}

\bibitem[]{}
Giavalisco, M., et al. 2004,
\textit{ApJ}, 600, L93

\bibitem[]{}
Haiman, Z. 2002,
\textit{ApJ}, 576, L1

\bibitem[]{}
Hu, E. M., Cowie, L. L., McMahon, R. G., Capak, P., Iwamuro, F.,
Kneib, J. -P., Maiharam T., \& Motohara, K. 2002,
\textit{ApJ}, 568, L75 (Erratum, 576, L99)

\bibitem[]{}
Hu, E. M., Cowie, L. L., Capak, P., McMahon, R. G., Hayashino, T., \&
Komiyama, Y. 2004,
\textit{AJ}, 127, 563

\bibitem[]{}
Jimenez, R., \& Haiman, Z. 2006,
\textit{Nature}, 441, 120

\bibitem[]{}
Kashikawa, N., et al. 2004,
\textit{PASJ}, 56, 1011

\bibitem[]{}
Kashikawa, N., et al. 2006,
\textit{ApJ}, 648, 7

\bibitem[]{}
Kawai, N., et al. 2006,
\textit{Nature}, 440, 184

\bibitem[]{}
Kodaira, K., N., et al. 2003,
\textit{PASJ}, 55, L17

\bibitem[]{}
Loeb, A., \& Barkana, R. 2001,
\textit{ARAA}, 39, 19

\bibitem[]{}
Malhotra, S., \& Rhoads, J. E. 2002,
\textit{ApJ}, 565, L71

\bibitem[]{}
Mobasher, B., et al.  2005,
\textit{ApJ}, 635, 832


\bibitem[]{}
Murayama, T., et al. 2007,
\textit{ApJS}, 172, 523

\bibitem[]{}
Nagao, T., et al. 2004,
\textit{ApJ}, 613, L9

\bibitem[]{}
Nagao, T., et al. 2005a,
\textit{AA}, 459, 85

\bibitem[]{}
Nagao, T., et al. 2005b,
\textit{ApJ}, 631, L5

\bibitem[]{}
Nagao, T., et al. 2007,
\textit{AA}, 468, 877

\bibitem[]{}
Nagao, T., et al. 2008,
\textit{ApJ}, submitted

\bibitem[]{}
Norman, M. L. 2008,
\textit{"Proceedings of First Stars III"} Eds. Brian W. O'Shea, Alexander Heger 
\& Tom Abel, AIP Conference Ser. (arXiv:0801.4924)

\bibitem[]{}
Oh, S. P., Haiman, Z., \& Rees,M. J. 2001,
\textit{ApJ}, 553, 73

\bibitem[]{}
Ota, K., et al 2008,
\textit{ApJ}, submitted (arXiv.0707.1561)

\bibitem[]{}
Ouchi, M., et al. 2004,
\textit{ApJ}, 611, 660

\bibitem[]{}
Ouchi, M., et al. 2005,
\textit{ApJ}, 62-, L1


\bibitem[]{}
Ouchi, M., et al. 2008,
\textit{ApJS}, in press (arXiv:0707.3161)


\bibitem[]{}
Partridge, R. B., \& Peebles, P. J. E. 1967,
\textit{ApJ}, 147, 868

\bibitem[]{}
Rhoads, J. E., \& Malhotra, S. 2001,
\textit{ApJ}, 563, L5

\bibitem[]{}
Rhoads, J. E., et al 2000,
\textit{ApJ}, 545, L85

\bibitem[]{}
Rhoads, J. E., et al 2003,
\textit{AJ}, 125, 1006

\bibitem[]{}
Rhoads, J. E., et al 2005,
\textit{ApJ}, 621, 582

\bibitem[]{}
Richard, J., et al. 2008,
\textit{AA}, in press (arXiv:astro-ph/0606134)

\bibitem[]{}
Santos, M. R., Ellis, R. S., Kneib, J. -P., Richard, J., \& Kuijken, L. 2004,
\textit{ApJ}, 606, 683

\bibitem[]{}
Scannapieco, E., Schneider, R., \& Ferrara, A. 2003,
\textit{ApJ}, 589, 35

\bibitem[]{}
Schaerer, D. 2002,
\textit{AA}, 382, 28

\bibitem[]{}
Schaerer, D. 2003,
\textit{AA}, 397, 527

\bibitem[]{}
Schneider, R., Salvaterra, R., Ferrara, A., \& Ciardi, B. 2006,
\textit{MNRAS}, 369, 825

\bibitem[]{}
Scoville, N., et al. 2007,
\textit{ApJS}, 172, 1

\bibitem[]{}
Sekiguchi, K., et al 2004,
\textit{AAS}, 205, 8105

\bibitem[]{}
Shapley, A. E., Steidel, C. C., Pettini, M., \& Adelberger, K. L. 2003,
\textit{ApJ}, 588, 65

\bibitem[]{}
Shimasaku, K., et al. 2003,
\textit{ApJ}, 586, L111

\bibitem[]{}
Shimasaku, K., et al. 2006,
\textit{PASJ}, 58, 313

\bibitem[]{}
Sonneborn, G. 2008
\textit{in this volume}

\bibitem[]{}
tanway, E. R., et al. 2007,
\textit{MNRAS}, 376, 727

\bibitem[]{}
Stark, D. P., Ellis, R. S., Richard, J., Kneib, J. -P., Smith, G. P.,
\& Santos, M. R. 2007,
\textit{ApJ}, 663, 10

\bibitem[]{}
Steidel, C. C., Adelberger, K. L., Giavalisco, M., Dickinson, M., \& Pettini, M. 1999,
\textit{ApJ}, 519, 1

\bibitem[]{}
Steidel, C. C., Adelberger, K. L., Shapley, A. E., Pettini, M., Dickinson, M., \&  Giavalisco, M.
2003,
\textit{ApJ}, 592, 728

\bibitem[]{}
Taniguchi, Y. 2001,
\textit{The Japan-German Workshop on Studies of Galaxies in the Young Universe with
New Generation Telescopes}, eidted by N. Arimoto \& W. Duschul (astro-ph/0301097)

\bibitem[]{}
Taniguchi, Y., Shioya, Y., Ajiki, M., Fujita, S. S., Nagao, T., \& Murayama, T. 2003,
\textit{JKAS}, 36, 123 (Erratum, 36, 283) 

\bibitem[]{}
Taniguchi, Y., et al. 2005,
\textit{PASJ}, 57, 165 (Erratum, 59, 277)

\bibitem[]{}
Taniguchi, Y., et al. 2008,
\textit{ApJ}, to be submitted


\bibitem[]{}
Tornatore, L., Ferrara, A., \& Schneider, R. 2007,
\textit{MNRAS}, 382, 945

\bibitem[]{}
Tumlinson, J., Giroux, M. L., \& Shull, M. 2001,
\textit{ApJ}, 550, L1

\bibitem[]{}
Tumlinson, J., Shull, M., \& Venkatesan, A. 2002,
\textit{ApJ}, 584, 608

\bibitem[]{}
Tumlinson, J., \& Shull, M. 2000,
\textit{ApJ}, 528, L65

\bibitem[]{}
Willis, J. P., \& Courbin, F. 2005,
\textit{MNRAS}, 357, 1348

\bibitem[]{}
Willis, J. P., Courbin, F., Kneib, J. -P., \& Minniti, D. 2008,
\textit{MNRAS}, in press (arXiv:0709.1761)

\bibitem[]{}
Yamada, F. S., et al. 2005
\textit{PASJ}, 57, 881

\bibitem[]{}
Yan, H., Dickinson, M. Giavalisco, M. Stern, D. Eisenhardt, P. R. M.; Ferguson, H. C. 2006
\textit{ApJ}, 651, 24


\end{thebibliography}
\end{document}